# Intelligent Exercise and Feedback System for Social Healthcare using LLMOps

Yeongrak Choi[1,*], Taeyoung Kim[2], Hyung Soo Han[1]

[1]Department of Biomedical Science, Graduate School, Kyungpook National University,

Daegu, Republic of Korea

[2]AIFactory, Daejeon, Republic of Korea

*Correspondence: ian.choi@knu.ac.kr

*Abstract* – This study addresses the growing demand for personalized feedback in healthcare platforms and social communities by introducing an LLMOps-based system for automated exercise analysis and personalized recommendations. Current healthcare platforms rely heavily on manual analysis and generic health advice, limiting user engagement and health promotion effectiveness. We developed a system that leverages Large Language Models (LLM) to automatically analyze user activity data from the "Ounwan" exercise recording community. The system integrates LLMOps with LLM APIs, containerized infrastructure, and CI/CD practices to efficiently process large-scale user activity data, identify patterns, and generate personalized recommendations. The architecture ensures scalability, reliability, and security for large-scale healthcare communities. Evaluation results demonstrate the system's effectiveness in three key metrics: exercise classification, duration prediction, and caloric expenditure estimation. This approach improves the efficiency of community management while providing more accurate and personalized feedback to users, addressing the limitations of traditional manual analysis methods.

## 1 Introduction

The digital healthcare market is experiencing unprecedented growth, projected to expand from $264.1 billion in 2023 to $1,190.4 billion by 2032, with a remarkable CAGR of 16.7% [1]. This explosive growth reflects the increasing adoption of digital platforms in healthcare, leading to a growing demand for personalized feedback and analysis-based recommendations. Healthcare platforms and social communities are rapidly embracing digital transformation across multiple channels, from telehealth solutions to wearable healthcare devices, with fitness and exercise tracking emerging as a key usage pattern, accounting for 47% of digital health applications. Also, healthcare platforms and social communities have increasingly adopted digital platforms to promote user engagement and health enhancement. Personalized recommendation systems play a vital role in enhancing user satisfaction and engagement across various platforms and services [2]. These systems analyze user behavior and preferences to provide tailored recommendations, significantly reducing the time users spend searching for relevant information or products [3].

Traditional healthcare feedback systems face limitations due to their reliance on manual processes and generalized advice, which fail to address individual user needs effectively. These constraints are further complicated by the complex integration requirements for diverse exercise devices, non-standardized data formats, and increased risk of data errors from manual processes. The large volume of user-generated data makes it challenging to provide timely and relevant feedback, while the difficulty in reflecting diverse user health conditions creates substantial personalization barriers.

The integration of advanced technologies with Large Language Models (LLMs) such as LLMOps offers a promising solution to these challenges [4]. By automating all the operations from data preparation to leveraging models with versioning, monitoring, evaluation, and continuous improvement, the exercise analysis process and delivering personalized recommendations are effectively streamlined as feedback loop while reducing manual intervention. Personalized recommendation methods, whether based on machine learning or deep learning, have become central to improving user experience in digital health platforms. These systems analyze user behavior and preferences to provide tailored recommendations, significantly reducing the time users spend searching for relevant information. The integration of LLMOps with LLM Framework, LLM APIs, Containerized infrastructure, and Continuous Integration/Deployment (CI/CD) enables efficient analysis of large amounts of user activity data, pattern identification, and tailored improvement recommendations.

The healthcare ecosystem's technological advancement is particularly evident in the integration of artificial intelligence and digital therapeutics. AI-powered systems enhance diagnostic accuracy and enable personalized treatment strategies, while digital therapeutics provide revolutionary approaches to prevention and treatment. This evolution supports the growing demand for data-driven, individualized experiences that enhance accessibility, transparency, and overall user satisfaction.

The importance of exercise analysis and feedback systems extends beyond mere tracking, playing roles in identifying areas for improvement, enhancing health outcomes, and sustaining participation. These systems provide motivating feedback that encourages progress and healthy competition while supporting heart health and chronic disease prevention. The integration of LLMOps into personalized recommendation systems marks a significant advancement in managing LLM-driven applications, enabling enterprises to enhance the efficiency and reliability of large-scale machine learning models.

As healthcare communities continue to grow in importance and the demand for personalized feedback increases, the need for innovative solutions becomes more pressing. The proposed system addresses these challenges by leveraging LLMOps to provide comprehensive analysis of user activity records. The system's architecture ensures scalability, reliability, and security, making it particularly suitable for large-scale healthcare communities while maintaining the continuous monitoring and optimization necessary for sustained accuracy and efficiency.

Exercise healthcare communities enable users to facilitate interaction and feedback among users by uploading their exercise records and providing comments to other users' exercise. This leads to improved health outcomes through shared experiences and mutual support. However, there are challenges of current healthcare community systems in analyzing and providing feedback.

First, the manual analysis and feedback process is highly time-consuming and inefficient, leading to substantial delays in user responses. These delays significantly impact on the effectiveness of healthcare interventions and result in diminished user engagement levels due to the extended feedback latency periods.

In terms of personalization, the system predominantly relies on generalized health advice rather than tailored recommendations. The absence of individualized recommendation systems has led to decreased user satisfaction levels. Users often receive generic feedback that fails to address their specific needs, resulting in limited adherence to health recommendations and reduced overall effectiveness of the healthcare initiatives.

The technical infrastructure presents another set of critical challenges. As the user base continues to grow, systems struggle to handle the overwhelming volume of data effectively. Traditional systems experience significant performance degradation, resulting in increased response times and delivery latency. These technical limitations create substantial scalability constraints that directly impact the quality of user experience.

Furthermore, integration complexities pose significant obstacles to system effectiveness. The challenges of integrating various technologies create operational inefficiencies, while maintaining security and reliability becomes increasingly difficult. These issues lead to elevated operational costs and, more importantly, diminished user trust due to system inconsistencies. The combination of these integration challenges significantly impacts the overall system reliability and user confidence in the platform.

This research investigates the application of LLMOps in exercise healthcare communities to enhance the efficiency and reliability of large-scale machine learning models while delivering personalized feedback and analysis of user activity records. The study specifically focuses on the "Ounwan" exercise record-sharing community, developing a platform that enables users to track, share, and receive feedback on their fitness activities through experimental implementations.

The research encompasses four primary areas of development and enhancement. First, the automated analysis system development focuses on implementing LLMOps for streamlined data processing, establishing real-time feedback generation mechanisms, and integrating various exercise data sources to create a comprehensive analytical framework. Second, the personalization enhancement initiative concentrates on developing sophisticated user-specific recommendation systems, implementing adaptive feedback mechanisms that evolve with user progress, and creating personalized exercise goals tailored to individual capabilities and preferences. Third, scalability optimization efforts center on designing robust distributed computing architecture, implementing efficient data processing systems, and enhancing overall system performance metrics to ensure consistent service delivery. Fourth, the platform integration component addresses the seamless incorporation of various technologies, implementation of comprehensive security measures, and development of maintenance optimization systems to ensure long-term sustainability.

This comprehensive objective aims to transform traditional manual-intensive processes into an automated, scalable system that effectively serves an expanding user base while maintaining exceptional performance and user satisfaction levels. The implementation within the Ounwan community demonstrates the potential of LLMOps to revolutionize healthcare community management, showcasing its ability to deliver personalized, efficient, and reliable exercise analysis and feedback at scale.

The organization of this paper is as follows. Section 2 reviews the related work in the field of healthcare informatics and exercise analysis, highlighting the limitations of existing systems. In Section 3, we introduce the exercise analysis system with Ounwan exercise community and discuss its components and functionalities. Section 4 explores the use of Large Language Model Operations (LLMOps) for providing personalized feedback to users. We describe the design and implementation details with LLMOps in Section 5, including the integration of LLMOps with LLM APIs, and Continuous Integration/Deployment (CI/CD). Section 6 presents the evaluation of the proposed system, including its accuracy and efficiency. Finally, we conclude the paper with a summary, contributions, and possible future work.

## 2   Related Work

The integration of IT technologies and artificial intelligence in healthcare, particularly through Large Language Models (LLMs), represents a significant advancement in medical technology and patient care delivery. This chapter examines the current landscape of healthcare applications, focusing on several key areas: the fundamental role of regular exercise in health maintenance, digital interventions in healthcare, analysis and feedback systems, and the emergence of AI-assisted healthcare solutions.

### 2.1  Regular Exercises for Health

Regular physical activity plays a fundamental role in maintaining overall health and well-being, with research demonstrating comprehensive benefits across multiple health domains. Studies have shown that consistent exercise can reduce premature death risk by 20-35%, highlighting its significance in longevity and quality of life [5].

In terms of physical health, regular exercise strengthens cardiovascular function, enhances muscular and skeletal health, and improves sleep quality through increased melatonin production [6]. The impact on disease prevention is equally significant, with research indicating that physical activity can reduce the risk of type-2 diabetes by 6% for every 500 kcal of weekly energy expenditure [7]. Moreover, physically active individuals demonstrate a 29% lower cancer-related mortality rate compared to sedentary populations [5].

Mental health benefits are also substantial, with exercise improving cognitive function, enhancing mood through endorphin release, and reducing symptoms of anxiety and depression. The increased blood flow and Brain-Derived Neurotrophic Factor (BDNF) production during exercise contribute significantly to brain health and cognitive function [6].

Functional benefits extend particularly to older adults, where regular physical activity improves musculoskeletal fitness, reduces fall risk, and maintains functional independence. Recent meta-analyses have demonstrated that moderate-intensity exercise, maintained for at least six weeks, significantly improves energy levels and reduces fatigue in healthy individuals [6]. Perhaps most striking is the finding that physically active individuals show a 52% lower risk of all-cause mortality compared to inactive individuals [5].

The relationship between physical activity and health benefits appears to be dose-dependent, with even modest improvements in physical fitness associated with significant health benefits [5]. This evidence strongly supports the integration of regular physical activity into daily life as an essential component of maintaining overall health and well-being.

## 2.2 Digital interventions as System

Digital health interventions have emerged as transformative tools in modern healthcare delivery, offering systematic approaches to enhance health outcomes and promote behavior change. The World Health Organization's Classification of digital interventions, services and applications in health (CDISAH) provides a standardized framework for categorizing these interventions, organizing them around three key axes: digital health interventions, health system challenges, and digital services and application types [8].

These digital interventions have demonstrated significant clinical effectiveness across various health conditions. In diabetes management, digital healthcare technologies have shown promising results in decreasing HbA1c levels compared to traditional care methods, with tailored mobile coaching proving particularly effective in improving glycemic control and reducing hospitalization risk for type 2 diabetes patients. Tailored mobile coaching (TMC) was effective in improving glycemic control and reducing the risk of hospitalization in individuals with type 2 diabetes [9]. Healthcare providers have also benefited from enhanced patient monitoring capabilities through these digital platforms to monitor patients closely [10].

Weight management programs utilizing digital interventions have achieved notable success rates, with studies showing that 75% of users maintained a 5% weight loss after one year, and nearly half maintained at least a 10% weight loss during the same period [11]. These results demonstrate the potential of digital interventions to support long-term behavior change and health improvement.

The WHO classification framework serves as a crucial resource for stakeholders across the health and technology sectors, including government agencies, healthcare providers, implementers, and researchers, providing a common language to articulate problems and needs that digital interventions can address. This standardized approach helps facilitate the development, implementation, and evaluation of digital health solutions while supporting inventory analysis, planning, and investment coordination in healthcare systems.

### 2.2.1 System Integration and Implementation Considerations

The integration of digital health interventions into existing healthcare systems presents complex challenges that require careful consideration and systematic approaches. According to Park et al. (2022), successful implementation demands seamless integration with existing healthcare infrastructure while maintaining operational efficiency [12]. This integration process involves establishing robust data exchange protocols between different healthcare systems, enabling real-time monitoring capabilities, and implementing standardized health data formats.

Research by Mitchell et al. (2023) demonstrates that effective system integration requires addressing three key components: infrastructure compatibility, data interoperability, and workflow optimization [13]. Healthcare organizations must ensure their existing systems can communicate effectively with new digital interventions while maintaining data integrity and security.

### 2.2.2 Implementation Challenges

The implementation of digital health interventions faces several significant challenges that can be categorized into three main areas:

- Technical Challenges: Network infrastructure limitations often present significant barriers to implementation, particularly in resource-constrained settings. Research indicates that system interoperability issues continue to plague healthcare organizations, with fragmented and unsustainable systems being a major challenge [14]. Data security, privacy concerns, and quality of digital health information remain paramount concerns.

- Healthcare Delivery Challenges: The integration of digital interventions into clinical workflows presents unique challenges. Studies show that healthcare professionals often resist changes to established workflows, particularly when new systems require significant adaptation of existing practices [13]. Limited digital health literacy among both healthcare providers and users necessitates comprehensive training programs and ongoing support systems.

- Evaluation Challenges: The rapid pace of technological advancement creates a mismatch with traditional clinical evaluation timeframes [15]. Organizations must strike a delicate balance between maintaining user experience and ensuring clinical efficacy while adhering to regulatory requirements. The evaluation process must also consider the continuous evolution of technology and its impact on healthcare delivery models.

These challenges underscore the importance of adopting a comprehensive approach to digital health intervention implementation, one that considers technical, organizational, and human factors while maintaining focus on improved healthcare outcomes and user experience [10].

2.2.3    Monitoring and Analysis System with Feedback Loop

The evolution of feedback loop architectures with monitoring and analysis in autonomic computing, from MAPE-K (Monitor-Analyze-Plan-Execute over shared Knowledge) to FOCALE (Foundation – Observe – Compare – Act – Learn – rEason), shares significant parallels with healthcare systems' analytical feedback mechanisms [17]. This relationship is evidenced by the COSARA intensive care platform, where the implementation of autonomic control loops achieved a 13.04% reduction in data execution time. Body Area Networks exemplify this connection through their continuous monitoring and automatic adaptation of vital signs such as heart rate, temperature, and ECG, operating similarly to the autonomic nervous system's unconscious regulation of vital bodily functions [17].

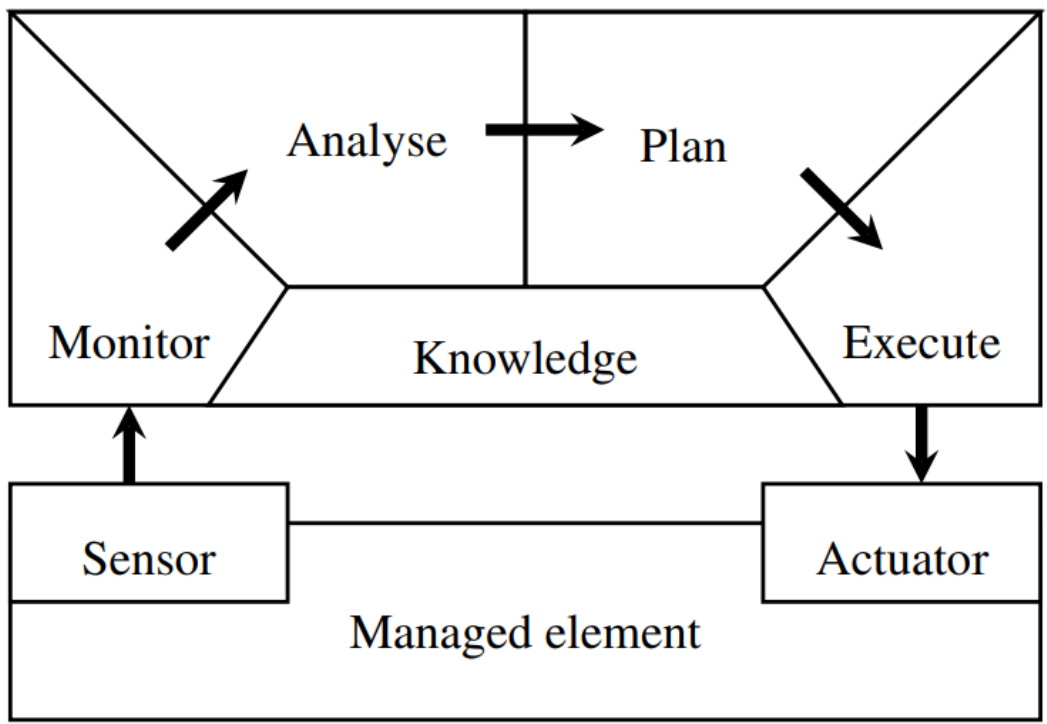

**Figure 1.** The MAPE-K autonomic feedback loop [16]

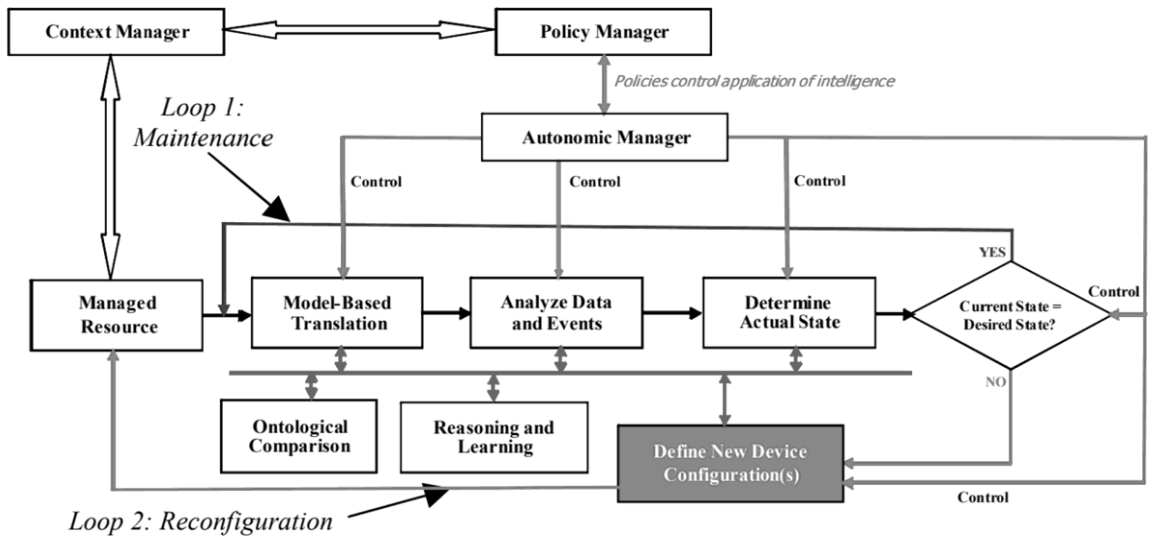

**Figure 2.** FOCALE autonomic architecture [19]

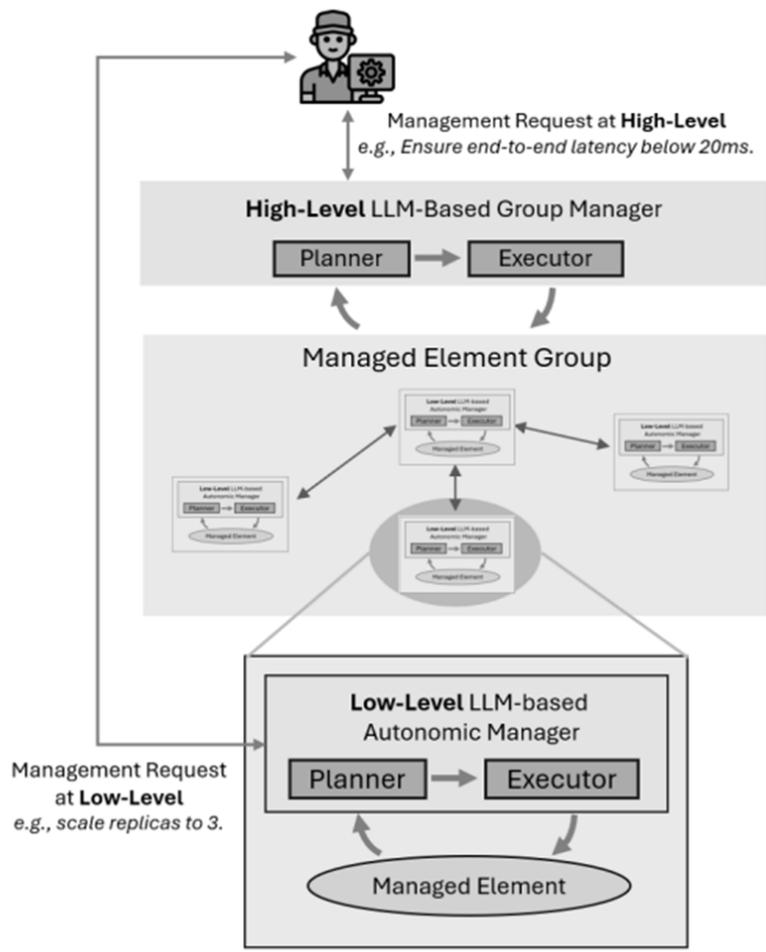

**Figure 3.** Autonomic computing with LLM-based multi-agent design [20]

Recent advancements in Digital Twin technology have further enhanced this integration by creating virtual representations of medical data and hospital environments, enabling real-time monitoring and predictive analytics [18]. These systems share core characteristics of autonomic computing, including self-configuration, self-healing, self-optimization, and self-protection.

The latest research in autonomic computing demonstrates promising developments through LLM-based implementations. The Vision of Autonomic Computing study shows that LLM-based multi-agent frameworks can achieve Level 3 autonomy in microservice management, effectively handling tasks such as proactive issue detection and basic self-healing capabilities [20]. This implementation simplifies the traditional MAPE-K loop into a more streamlined Plan-Execute feedback mechanism, where LLMs handle both the planning and execution phases of system management.

The convergence of healthcare feedback systems and autonomic computing principles, now enhanced by LLM capabilities, represents a significant step forward in developing truly self-managing systems that can adapt and respond to changing conditions autonomously [17][20].

### 2.2.4 AI-assisted Analysis and Feedback

Recent research demonstrates advancements in AI-assisted analysis and feedback systems in healthcare. Interactive analysis and feedback systems have emerged as a crucial component, with studies highlighting the importance of enhancing human-AI interaction efficiency and effectiveness in healthcare settings [21].

A practical implementation of this concept is demonstrated through an AI chatbot system built on a Kubernetes-based scalable architecture. This system leverages BERT and LSTM models to recommend appropriate medical specialties based on patients' natural language symptom descriptions, showcasing the potential of AI in improving patient triage and care pathways [22].

In clinical applications, an AI-based sepsis early warning system has shown remarkable results, achieving a 10% improvement in sepsis bundle compliance and a 17% reduction in in-hospital sepsis-related mortality. This implementation emphasizes the critical importance of continuous monitoring systems in maintaining adaptability and improving patient outcomes [23].

Recent developments in LLM operations (LLMOps) with CI/CD pipelines have further advanced healthcare applications, offering enhanced security measures, flexible deployment options, and maintained accuracy in meeting medical requirements [24]. This approach demonstrates how modern AI infrastructure can be effectively integrated into healthcare systems while maintaining high standards of reliability and security.

These research findings collectively demonstrate that AI applications in healthcare extend beyond mere technological implementation, contributing to tangible improvements in clinical outcomes and patient care. The consistent emphasis across studies on continuous monitoring and system improvement highlights the evolutionary nature of AI integration in healthcare settings.

# 3 Exercise Analysis System with Community

The Intelligent Exercise Analysis and Feedback System is designed to integrate seamlessly with the Ounwan exercise healthcare community platform, where users actively share and interact with exercise-related content. This social fitness platform serves as a hub for users to document their fitness journey, share achievements, and engage with fellow fitness enthusiasts.

## 3.1 Community Platform

The exercise community platform facilitates various user interactions centered around fitness activities. Users can upload their exercise records, including workout details, duration, intensity, and personal achievements. These posts become focal points for community engagement, where members can view, comment, and interact with each other's content, fostering a supportive environment for fitness motivation.

As of November 2024, there are 133 members joining in the community, and users post exercise results with exercise result screenshots and photos.

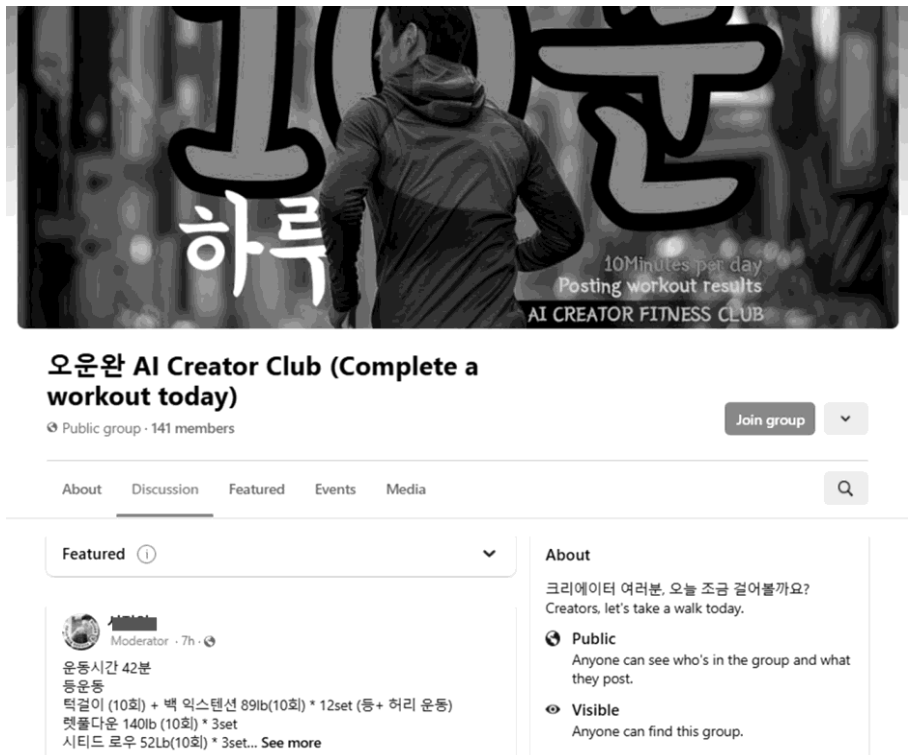

**Figure 4.** Ounwan exercise healthcare community - screenshot

## 3.2 Analysis and Feedback System

The system enhances the Ounwan community by implementing automated analysis and feedback generation through LLMOps integration. The system use diagram, as illustrated in Figure 5, shows the interactions between users, the Ounwan exercise community platform, and the intelligent exercise analysis & feedback system. The system conducts data collection operations with periods such as daily, systematically gathering exercise-related information from the community. Through its exercise datafication process, it transforms unstructured workout data into standardized, analyzable formats. The system then employs sophisticated analysis algorithms to process this structured data, deriving meaningful insights about user performance and patterns.

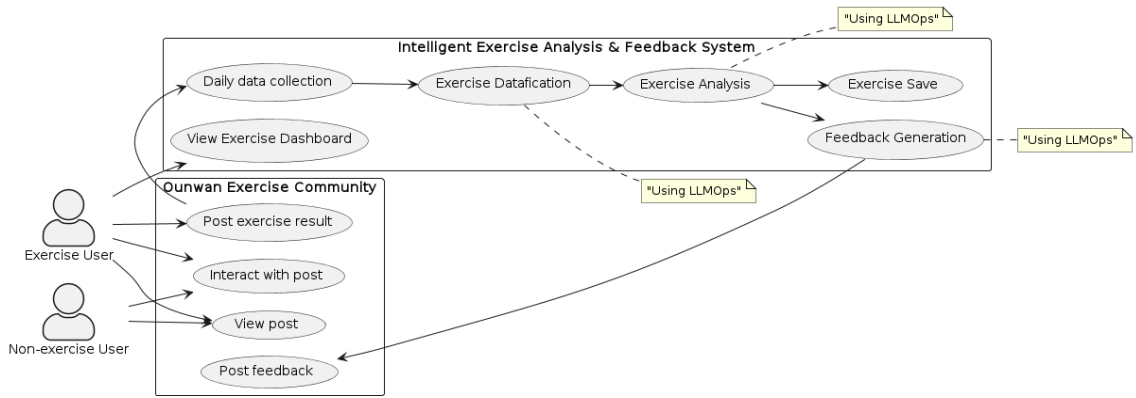

**Figure 5.** Use case diagram for the system with community

The analyzed data is securely stored in a dedicated database for future reference and longitudinal analysis. Finally, the system generates personalized feedback based on the comprehensive analysis of user data, providing valuable insights and recommendations to community members. This automated analysis and feedback loop creates a continuous cycle of improvement and engagement within the exercise community.

The integration between the community platform and the analysis system creates a robust ecosystem that promotes user engagement while delivering data-driven, personalized exercise guidance. This systematic approach ensures that users receive timely, relevant feedback while maintaining an engaging social fitness environment.

## 3.3 Modular Architecture

The system implements a scalable, container-based modular architecture that efficiently manages exercise analysis and feedback generation. This architecture is designed to handle growing user demands while maintaining system performance and reliability.

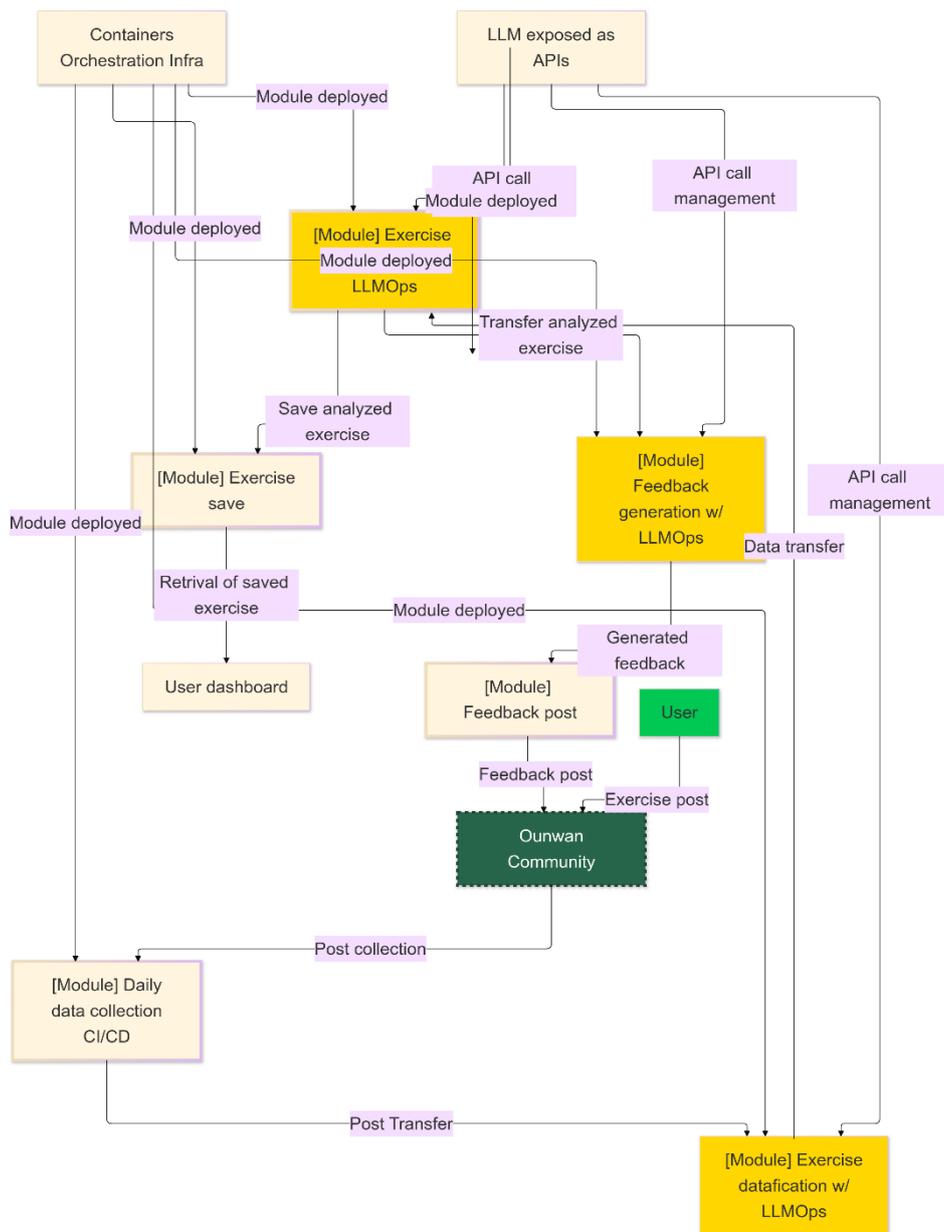

**Figure 6.** System modular architecture

### 3.3.1 Container-Based Infrastructure

The system's core functionality is distributed across several containerized modules deployed using orchestration infrastructure. The infrastructure deploys the target modules with scalability management. In the proposed system, all the target modules are containerized and deployed into the orchestration infrastructure. These modules include daily data collection for exercise records, exercise analysis processing, exercise data storage, and feedback generation components. The containerized approach ensures system scalability and efficient resource utilization while maintaining isolation between different system components.

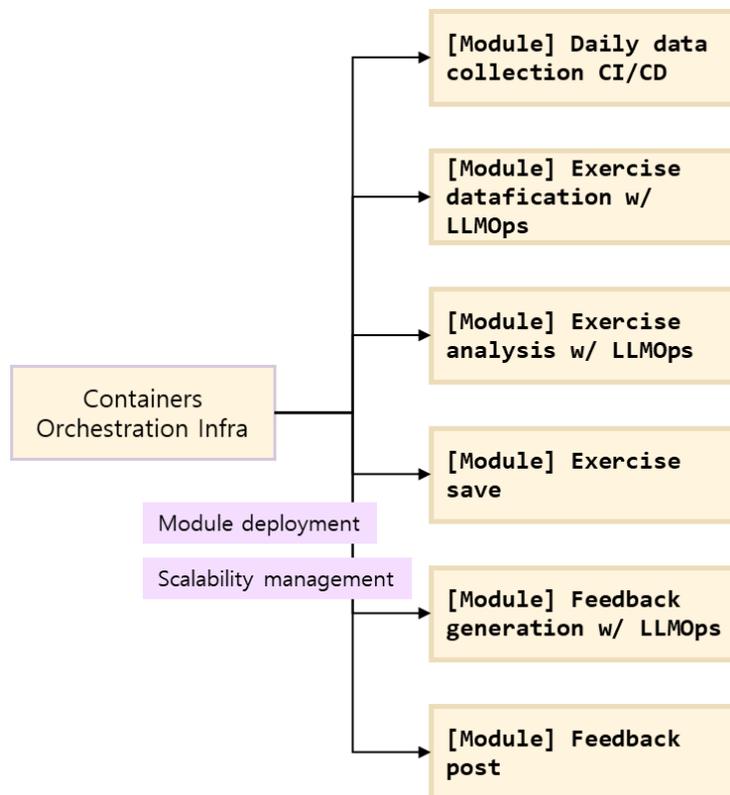

**Figure 7.** Containerized module deployment infrastructure

Table 1 shows an example of a Dockerfile for the daily data collection module. The Dockerfile starts by pulling a lightweight base image to ensure a minimal and efficient container environment. The system dependencies are installed through the package manager. For daily data collection module, required Chromium browser, Chromium WebDriver, and Noto CJK fonts for Korean language support packages are installed. For the programming language dependencies, Python packages are installed using pip. In the thesis, the data collection module was implemented with Selenium version 4.25.0 being a key requirement for web automation functionality, and additional dependencies are managed through a requirements.txt file to maintain a clean and organized dependency structure. The application setup involves copying essential files such as app.py, environment configuration (.env), and the data file (posts.txt) into the container. Finally, the container's entry point is configured with a CMD instruction that specifies the Python interpreter to execute the main application file, ensuring proper initialization of the application when the container starts.

**Table 1.** Example of Dockerfile for a module

```
# Base Image for the target container module
FROM python:3.9-slim

# Install essential System Dependencies
RUN dnf install -y \
    chromium \
    chromium-driver \
    fonts-noto-cjk

# Python Dependencies
RUN pip install selenium==4.25.0 \
    && pip install -r requirements.txt

# Application Setup
COPY app.py .env ./
COPY posts.txt ./

CMD ["python", "app.py"]
```

The system's infrastructure leverages enabling dynamic execution of system modules based on specific triggers and scheduled events. The encapsulation strategy with container technologies maintains consistent performance across different deployment environments by effectively managing dependencies and runtime configurations.

### 3.3.2    Applying LLMOps into modular architecture

At the core of the system, the Exercise module with LLMOps serves as a central processing unit that interfaces with LLM APIs exposed through standardized endpoints. This module handles the complex task of exercise analysis while maintaining efficient API call management. The system implements a sophisticated data flow where exercise information is collected through the Daily Data Collection CI/CD module, which regularly gathers posts from the Ounwan Community platform.

The Exercise Datafication module transforms raw exercise data into structured formats suitable for analysis. Once processed, the analyzed exercise data is stored through the Exercise Save module, which maintains a repository of user exercise information. This stored data is then accessible through the user dashboard, providing a comprehensive view of exercise patterns and progress.

The Exercise Analysis module leverages LLMOps capabilities through the AIModelComparator, as shown in Table 2. The AIModelComparator initializes LLM models and processes queries by combining specified prompts with text and images for each model. This parallel processing approach allows simultaneous execution of multiple AI models, enhancing operational efficiency. Then, data processing is conducted according to the Data Processing Logic outlined in Table 3. This logic implements batch processing for memory optimization while maintaining error handling at each stage. The system processes entries in parallel batches, with each entry including post information and associated images. The architecture tracks key performance metrics such as execution time, token usage, and cost calculations, allowing comprehensive monitoring and observability. The module's architecture supports scalability, enabling seamless integration of new AI models. This flexibility is achieved through a modular structure that standardizes the interface for model integration while maintaining consistent error handling and performance tracking across all models. Parallel execution enhances processing efficiency and enables real-time comparative analysis of different model performances.

**Table 2.** Exercise Analysis: *AIModelComparator* (pseudo-code)

```
Class AIModelComparator:
    Initialize:
        Set up LLM models
        Configure system prompt for exercise analysis

    Function encode_image(image_path):
        Open image
        Convert to base64 string
        Return encoded image

    Function prepare_messages(text, image_paths):
        Create message list with system prompt
        If images exist:
            Encode each image
            Add text and images to message
        Else:
            Add only text to message
        Return messages

    Async Function analyze_with_model(model, text, image_paths, model_name):
        Start timer
```

```
        Prepare messages
        Try:
            Get model response with token tracking
            Calculate execution time
            Return results and metrics
        Catch:
            Return error information

    Async Function compare_models(text, image_paths):
        Execute analysis on all models in parallel
        Return combined results
```

**Table 3.** Exercise Analysis: *Data Processing Logic* (pseudo-code)

```
Async Function process_entry(comparator, entry):
    Extract post information
    Format analysis request text
    Prepare image paths
    Try:
        Get results from all models
        Parse JSON responses
        Return structured output
    Catch:
        Return error information

Async Function process_batch(comparator, entries, output_file):
    For each entry in batch:
        Process entry in parallel
        Write results to output file

Main Process:
    Initialize AIModelComparator
    Open input data file
    Create output file with timestamp
    For each batch in data:
        Process batch of entries
```

> Write results to output file
> Close output file

The Feedback Generation module, also powered by LLMOps, creates personalized feedback based on the analyzed exercise data. This feedback is then posted back to the Ounwan Community through the Feedback Post module, completing the feedback loop. The entire system maintains data integrity and flows through carefully managed API calls and data transfer protocols, ensuring efficient communication between all components while maintaining system reliability and scalability.

This architecture demonstrates a thoughtful integration of containerization, LLMOps, and API management, creating a robust system capable of handling complex exercise analysis and feedback generation tasks while maintaining high performance and user satisfaction.

## 3.4 Functions integrated with LLMOps

The system implements a scalable, container-based modular architecture that efficiently manages exercise analysis and feedback generation. This architecture is designed to handle growing user demands while maintaining system performance and reliability.

### 3.4.1 Exercise datafication

The system periodically collects exercise posts from the Ounwan community platform, transforming them into structured data using LLM framework. This process involves:

- Regular scanning and collection of community posts
- Extraction of exercise-related information from posts
- Conversion of unstructured data into standardized formats
- Storage of processed data for analysis

### 3.4.2 Exercise Analysis

The analysis component leverages LLM frameworks to process the digitalized exercise data:

- Pattern recognition in exercise behaviors
- Performance metric calculation
- Progress tracking across multiple dimensions
- Identification of potential areas for improvement
- Storage of analyzed results in a structured database

3.4.3　Feedback Generation

The system generates personalized feedback using LLM-powered analysis:

- Creation of context-aware exercise recommendations
- Development of personalized improvement suggestions
- Generation of motivational messages
- Automatic posting of feedback to the community platform

The LLMOps framework maintains continuous performance monitoring of the deployed LLM models, ensuring optimal operation throughout the system. This monitoring system tracks real-time performance metrics and conducts regular assessments of feedback accuracy, enabling immediate response to any quality variations. The implementation includes robust version control mechanisms for LLM models, facilitating systematic updates while maintaining system stability.

The deployment process is fully automated through container orchestration infrastructure, allowing for dynamic scaling based on system demands. This containerized approach ensures efficient resource utilization and system reliability. Quality assurance measures are embedded throughout the content generation pipeline, with API call management systems monitoring usage patterns and enforcing appropriate limits to maintain consistent performance. Additionally, the integration implementation includes error handling mechanisms and performance monitoring capabilities, ensuring system stability and providing insights into operational metrics. This approach to data processing and model management reflects LLMOps practices, combining efficiency, reliability, and scalability in one system.

Through this comprehensive LLMOps integration, the system delivers reliable, high-quality exercise analysis and personalized feedback while maintaining operational efficiency and scalability. The architecture's modular design allows for seamless updates and improvements while ensuring consistent service delivery to the Ounwan community platform.

# 4　Applying LLM for Analysis and Feedback

The integration of LLMOps into personalized recommendation systems marks a significant advancement in managing LLM-driven applications. LLMOps can provide accurate and personalized feedback to users by leveraging Large Language Models (LLMs) to analyze user activity records and provide tailored recommendations.

## 4.1 Exercise datafication

The exercise datafication process implements data collection and transformation approach, specifically tailored to handle the unique challenges of extracting information from the Ounwan community platform. Since the platform does not provide direct API access for individual posts, the system employs an automated web crawling mechanism to gather exercise-related data.

4.1.1    Data Collection Process

The system utilizes a scheduled web crawling mechanism that systematically accesses the Ounwan community platform through a web browser interface. This automated process performs daily scrolling operations to capture all posts published within the specified timeframe. Once the post listings are identified, the system extracts individual post URLs and processes each post to collect six essential data fields: *post_date, actor_id, photos, name, post_id,* and *content*.

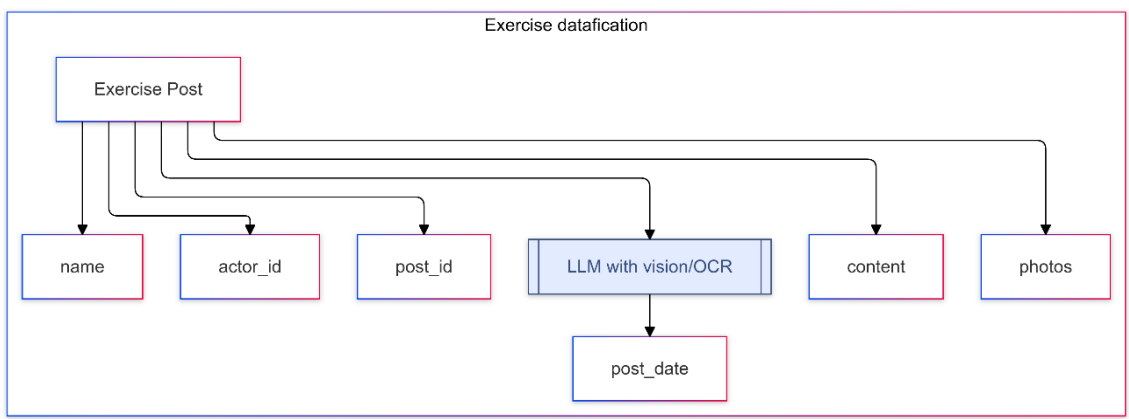

**Figure 8.** Exercise datafication and using LLM to overcome obfuscation

4.1.2    HTML Analysis and Data Extraction

The data extraction process involves detailed HTML structure analysis to precisely locate and retrieve specific data elements. The system employs advanced crawling techniques that identify and map the location of each data element within the HTML structure. This systematic approach ensures accurate data extraction while maintaining the integrity of the collected information.

4.1.3    Date Extraction with LLM to overcome obfuscation

A particular challenge arose with date extraction due to the platform's implementation of obfuscation techniques that prevent direct access to date information through conventional HTML parsing. Each character on HTML date field is sparse, as illustrated in Figure 9. To overcome this limitation, the system implements an innovative solution:

- Automated screenshot capture of individual posts by cropping date part

- Implementation of OCR (Optical Character Recognition) technology with LLM

- Specialized processing to extract and validate date information from the captured images

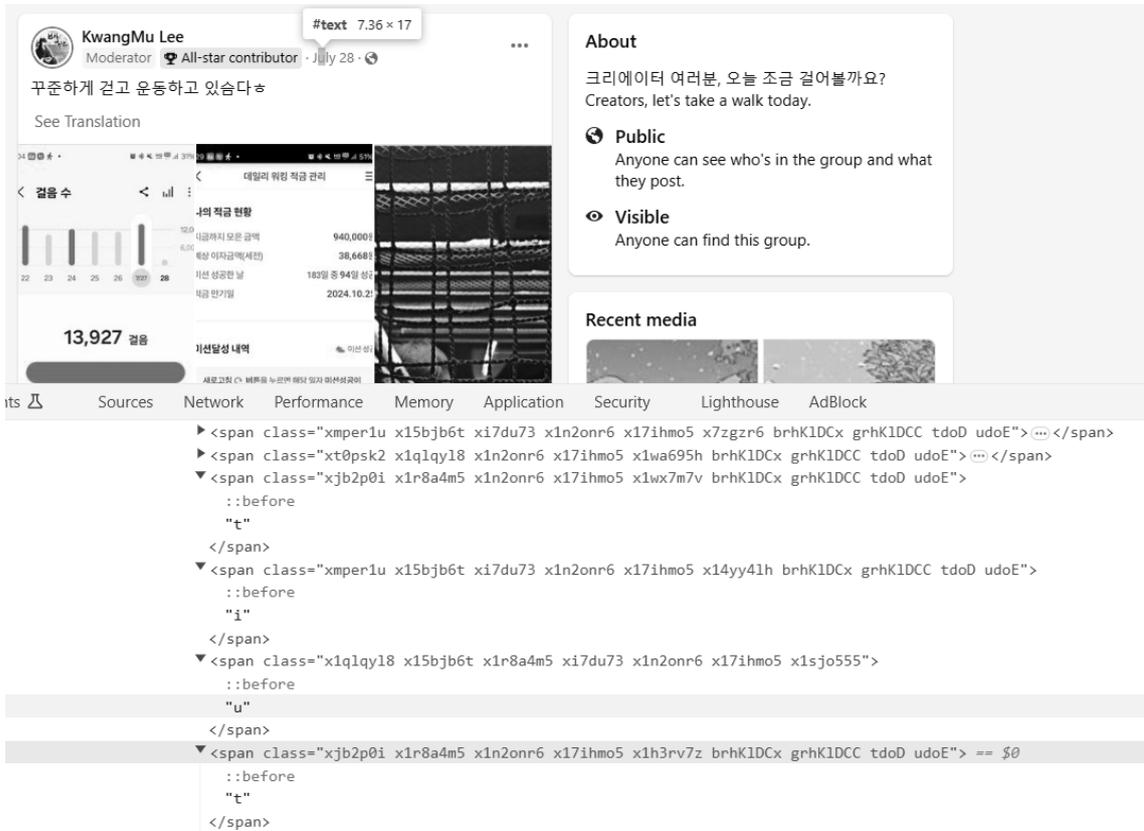

**Figure 9.** obfuscated date information on Ounwan community post

### 4.1.4 Data Transformation Pipeline

The collected data undergoes a transformation process through several stages. Initially, cleaning and validation are performed on all extracted content to ensure data integrity. Data formats across all fields are then standardized to maintain consistency throughout the system. For example, all date fields are transformed into the "YYYY-mm-dd" format, where YYYY is the four-digit year number, mm is the two-digit month number, and dd is the two-digit day number. Following this, OCR-extracted date information is integrated with other post data using LLM technology. Finally, the data is structured to ensure compatibility with subsequent analysis modules, preparing it for further processing and examination.

Figure 10 illustrates an example of Ounwan community exercise post, and datafication result with JSON notation. This datafication approach ensures reliable and consistent data collection while effectively addressing the platform's technical constraints through innovative solutions.

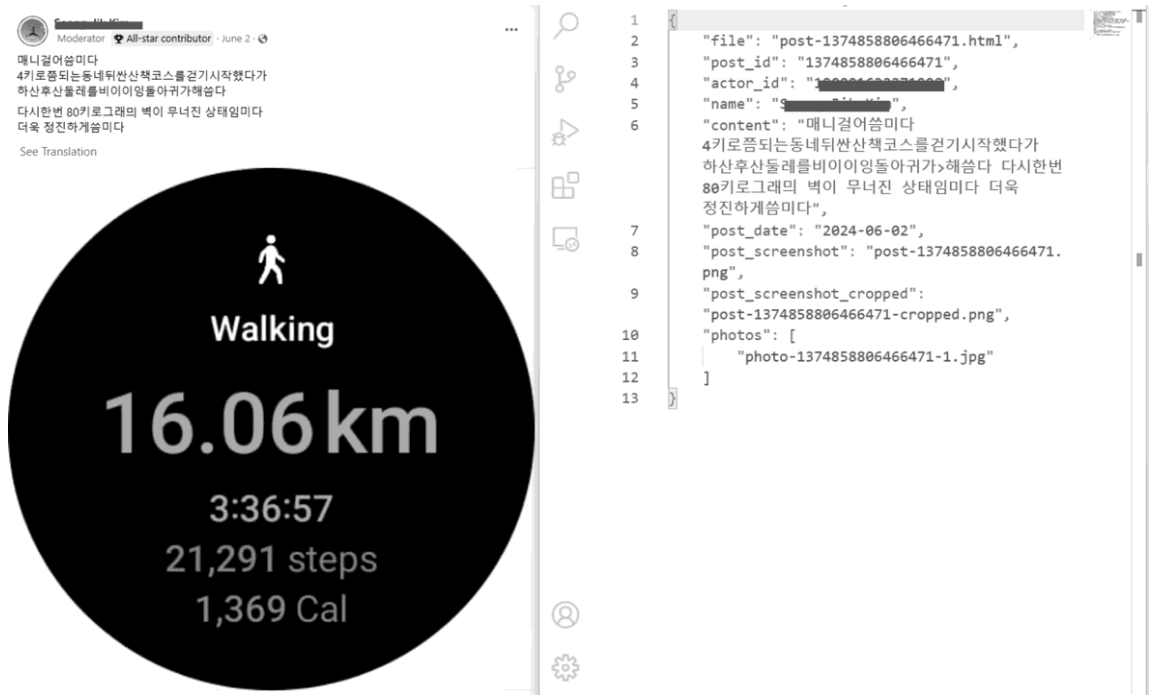

**Figure 10.** Example of Ounwan community exercise post and datafication result

## 4.2 Exercise analysis

The exercise analysis component processes structured exercise data through LLM-powered analytical framework. To standardize heterogeneous exercise data from various sources, the system focuses on three fundamental metrics: exercise verification (*is_exercise*), exercise duration (*exercise_duration*), and caloric expenditure (*calories*).

### 4.2.1 Core Analysis Framework

The system employs advanced Natural Language Processing (NLP) techniques to extract meaningful exercise information from user posts. This analysis pipeline processes structured data from the datafication module to identify and quantify key exercise parameters. The LLMOps integration enables pattern recognition and contextual understanding of exercise-related content.

### 4.2.2 Metric Analysis Components

The analysis system processes three standardized metrics that enable evaluation across different types of exercises:

- **Exercise Verification**: The type is for employing contextual analysis to validate whether a post contains legitimate exercise activity. This verification process examines post content to distinguish genuine exercise activities from non-exercise related content, ensuring data quality and reliability.

- **Duration Assessment**: The type is for calculating total exercise duration by analyzing temporal indicators within posts. This metric is fundamental for quantifying physical activity levels and serves as a key measure of exercise engagement. The duration tracking incorporates both explicit time statements and implicit temporal markers to generate standardized duration measurements.

- **Caloric Computation**: The system estimates energy expenditure by analyzing the exercise type and duration data. While direct calorimetry provides the most accurate measurements, the system employs validated algorithmic models to estimate caloric burn based on exercise intensity levels and duration. This computation accounts for different exercise modalities and their varying energy demands to provide standardized caloric expenditure metrics. These three metrics work together to create a comprehensive yet standardized way to analyze and compare different types of exercise activities, enabling consistent evaluation across various workout modalities. The standardization allows for meaningful comparisons and tracking of exercise patterns over time, regardless of the specific type of physical activity being performed.

### 4.2.3 Named Entity Recognition (NER) Implementation with LLM integration

The system utilizes specialized NER models powered by LLM to process exercise-related content within user posts through multiple stages. The detection process begins with tokenization and preprocessing of the text, where exercise terminology and expressions are identified through contextual analysis. The system performs entity classification to categorize different types of exercises and their associated intensity levels, utilizing both rule-based and machine learning approaches.

The temporal and quantitative information extraction involves analyzing specific patterns in the text to identify duration, sets, repetitions, and other numerical data. The system employs contextual analysis to understand the relationships between identified entities, ensuring accurate interpretation of exercise-related information within varying contexts. This comprehensive approach enables the system to handle complex exercise descriptions while maintaining high accuracy in entity recognition.

The analyzed data serves as the foundation for generating personalized feedback and recommendations, enabling the system to provide tailored insights based on each user's unique exercise patterns and preferences. This comprehensive analysis approach ensures accurate and consistent exercise assessment while maintaining adaptability to various exercise types and user behaviors.

## 4.3 Feedback generation

The feedback generation system employs LLM technology to create highly personalized exercise recommendations by analyzing both current and historical exercise data, and LLMOps for feedback analysis. This approach ensures that users receive contextually relevant and actionable feedback that supports their fitness journey.

### 4.3.1 Prompt Engineering Implementation

The system leverages prompt engineering techniques to generate optimized LLM feedback through a sophisticated multi-layered approach. At its core, the system employs a specialized system message that defines the LLM's role as a professional fitness trainer dedicated to analyzing exercise data and providing feedback. This role definition includes specific instructions about tasks to be performed and objectives to be met, ensuring consistent output formatting and response patterns like True/False, number-only, output in minutes, and specifying as JSON output. The prompts are dynamically constructed using instruction-based and directional stimulus techniques to incorporate individual exercise patterns and historical performance data. The system message framework ensures that the LLM maintains its persona as a knowledgeable fitness trainer while delivering:

- Consistent analysis of exercise patterns and performance metrics
- Standardized feedback format across different exercise types
- Professional-grade recommendations aligned with fitness industry standards
- Appropriate motivational messaging that maintains trainer-client dynamics

This structured approach to prompt engineering ensures that the LLM consistently delivers high-quality, professional-grade feedback while maintaining the role of a knowledgeable fitness trainer throughout all interactions.

### 4.3.2  Feedback Framework

The feedback system generates multi-dimensional recommendations that extend beyond simple encouragement:

- **Performance Analysis**: The system conducts detailed analysis of exercise patterns using multidimensional physical activity profiling. This approach enables more accurate depiction of physical activity that reduces misclassification while providing a holistic representation of user progress. The analysis identifies patterns in exercise behaviors and calculates various performance metrics across multiple dimensions.

- **Actionable Recommendations**: Based on analyzed data, the system generates personalized coaching recommendations that adapt to individual needs. These include progressive intensity adjustments based on performance data, complementary exercise suggestions derived from pattern analysis, recovery period recommendations using biometric feedback loops, and adaptation strategies informed by user capability assessments.

### 4.3.3  Engagement Enhancement

The feedback system is designed to maintain long-term user engagement through:

- Regular progress updates and milestone celebrations
- Personalized goal-setting recommendations
- Adaptive feedback based on user response patterns
- Motivational support during challenging periods

### 4.3.4  Integration with Exercise Analysis

The feedback generation module seamlessly integrates with the exercise analysis system by:

- Utilizing analyzed exercise metrics (is_exercise, exercise_duration, calories)
- Incorporating historical performance trends
- Considering user interaction patterns
- Adopting recommendations based on exercise consistency

This feedback system creates a dynamic, personalized experience that supports users in achieving their fitness goals while maintaining long-term engagement with the platform.

## 5 Implementation with LLMOps

The implementation of our exercise analysis and feedback system leverages LLMOps to ensure efficient operation, monitoring, and deployment of LLM-powered components. This section details the implementation approach, focusing on continuous integration/continuous deployment (CI/CD) practices and LLM performance monitoring.

### 5.1 CI/CD Implementation

The CI/CD implementation brings several key advantages to our exercise analysis and feedback system. Continuous Integration ensures consistent quality across codebases while reducing the likelihood of bugs in production. The automated testing and deployment processes minimize human error and ensure consistency across environments. The CI/CD pipeline is configured to build, test, and deploy the application within a containerized environment.

Below is a representative YAML declaration code for GitHub Actions workflow configuration as a sample to run daily at a specified time using cron [25] syntax:

**Table 4.** CI/CD - YAML declaration sample code for LLMOps

```yaml
name: CI/CD Pipeline

on:
  push:
    branches:
      - main
  pull_request:
    branches:
      - main

jobs:
  build:
```

```
      runs-on: ubuntu-latest
      steps:
      - name: Checkout code
        uses: actions/checkout@v2
      - name: Set up Docker Buildx
        uses: docker/setup-buildx-action@v1
      - name: Build and push Docker image
        uses: docker/build-push-action@v2
        with:
          context: .
          push: true
          tags: ountan/repository_image:latest

  deploy:
    runs-on: ubuntu-latest
    needs: build
    steps:
    - name: Deploy to production
      run: |
        echo "Deploying Docker image to production..."
```

This CI/CD implementation enables faster build times and more frequent releases, allowing the system to deliver updates and improvements to users more rapidly. The Continuous Deployment aspect is integrated with LLMOps to automatically deploy model updates and improvements based on performance monitoring data. This integration enables automated model transitioning, deployment, and monitoring while maintaining high reliability standards.

## 5.2 LLMOps Integration with LLM Framework

The LLMOps part offers various LLM integrations and tools for LLMs and operations for running and monitoring with LLMs. There are several open-source LLMOps frameworks. LangChain [26] provides a comprehensive framework for LLM application development. LangGraph [27], built on top of LangChain, excels at structuring complex, stateful multi-agent applications with features like cycles and state persistence. CrewAI [28] focuses on orchestrating role-playing autonomous agents for collaborative intelligence. LlamaIndex [29] specializes in data integration and vector store capabilities, making it particularly effective for sophisticated vector search and RAG implementations. Semantic Kernel [30] offers a lightweight approach with plugins and functions, emphasizing modular architecture for domain-specific applications. The thesis incorporated LangChain for implementation to create a robust and efficient development environment. This implementation facilitates seamless interaction with various LLM APIs while maintaining comprehensive monitoring and optimization capabilities across the entire system architecture.

The system utilizes LangChain, LLM Framework component to streamline LLM operations, enabling efficient prompt management and API interactions. This framework provides a standardized interface for multiple LLM providers, allowing for flexible model selection and optimization based on specific task requirements. The implementation includes comprehensive API usage monitoring through dedicated tracking mechanisms that observe request patterns, response times, and error rates.

Our monitoring system maintains detailed metrics on API consumption patterns, including:

- Request frequency and volume across different endpoints
- Response latency and performance metrics
- Token usage and cost optimization
- Error rates and failure patterns
- Model performance and accuracy metrics

The API management layer implements sophisticated rate limiting and load balancing mechanisms to ensure optimal resource utilization while preventing API quota exhaustion. This includes automated alerts for approaching usage limits and dynamic routing capabilities to maintain system availability during high-demand periods.

The containerized architecture facilitates efficient scaling and deployment of LLM-powered modules, with each component maintained in isolated containers for optimal resource management. The system's modular design allows for independent scaling of different components based on demand, ensuring efficient resource utilization while maintaining system performance.

Through this comprehensive integration of LLM frameworks and monitoring capabilities, the system maintains high reliability and performance standards while optimizing resource usage and cost efficiency. The implementation enables continuous improvement through detailed performance analytics and automated optimization processes, ensuring sustainable operation of the exercise analysis and feedback system.

# 6 Evaluation

The proposed system is evaluated based on its accuracy and efficiency. The evaluation is conducted using a combination of quantitative and qualitative methods.

## 6.1 Ounwan exercise community data

The evaluation of our exercise analysis system was conducted using a comprehensive dataset collected from the Ounwan exercise community platform over a specific time period. The dataset encompasses 741 posts uploaded between May 28, 2024, and November 5, 2024, representing the posts which usually have exercise activities and interactions of 133 active community members. Table 5 illustrates the overall data for evaluation.

Table 5. Summary of Ounwan exercise community data for evaluation

| Start date | End date | Number of members | Number of Posts |
|---|---|---|---|
| May 28, 2024 | Nov 5, 2024 | 133 | 741 |

To establish a reliable baseline for evaluating the LLMOps-powered exercise analysis system, we implemented a manual annotation process. This process involved creating ground truth data by individually reviewing and labeling each post to determine whether it contained legitimate exercise activities. The manual annotation process was meticulously conducted to ensure accuracy in the ground truth dataset, enabling precise evaluation of the system's exercise classification capabilities.

Table 6. Collected data distribution (daily)

| Number of posts daily | Minimum | Maximum | Average | Median |
|---|---|---|---|---|
| | 1 | 13 | 4.737 | 4 |

The dataset represents a diverse range of exercise activities and posting patterns, providing a robust foundation for assessing the system's ability to accurately identify and analyze exercise-related content. This comprehensive ground truth dataset serves as a benchmark for evaluating the performance of our LLMOps-based analysis system, particularly in distinguishing exercise-related posts from general community interactions.

Table 7. Ground truth data for evaluation

| is_exercise: True | duration | | calories | |
|---|---|---|---|---|
| 712 | Count: 474 | Average: 96.76 | Count: 476 | Average: 563.17 |

The temporal span of approximately five months ensures that the evaluation captures various patterns in user behavior and exercise reporting, making it a reliable indicator of the system's real-world performance. This evaluation approach enables us to assess both the accuracy and consistency of our automated exercise analysis system across different types of exercise posts and user interactions.

6.2 Evaluation Methodology

The evaluation methodology for our exercise analysis system implements a comprehensive approach to assess the accuracy of LLM-based exercise analysis against manually labeled ground truth data. The evaluation process encompasses multiple dimensions of exercise analysis, focusing on both qualitative and quantitative aspects of exercise identification and measurement.

6.2.1    Data Processing Approach

The evaluation process begins with the application of LLM to analyze exercise data from two primary sources: textual content from user posts and accompanying screenshot images. This dual-source analysis enables a more robust understanding of exercise activities, as the system processes both explicit textual descriptions and visual evidence of exercise completion. The LLM framework processes this information using standardized prompts to ensure consistency across all analyzed posts.

6.2.2    Analysis Parameters

The system evaluates exercise data across multiple dimensions:

- Exercise identification (*is_exercise*)

- Exercise duration quantification

- Caloric expenditure estimation

Each parameter is analyzed using carefully crafted prompts that guide the LLM in extracting relevant information while maintaining consistency with the ground truth data format.

6.2.3    Accuracy Assessment

The evaluation methodology implements a systematic comparison between LLM-generated analysis results and manually labeled ground truth data. This comparison process involves:

- Direct matching of exercise classification results

- Verification of exercise duration calculations

- Validation of caloric expenditure estimates

The system maintains consistent units and measurement standards across both automated analysis and ground truth data to ensure accurate comparison. This standardization is crucial for calculating meaningful accuracy metrics that reflect the system's real-world performance.

6.2.4    Validation Process

The validation process is enhanced through the integration of LLMOps with LangChain framework and CI/CD systems using GitHub Actions, enabling comprehensive testing across multiple LLM providers. Our system implements automated evaluation pipelines that assess the performance of three leading LLM models: OpenAI's GPT model, GPT model from Azure OpenAI services, and Amazon Bedrock's Claude 3. For temperature, the experiment used default value: 0.7.

**Table 8**. LLM models used for evaluation

| Model Provider | Model name | Model version | Context Window |
|---|---|---|---|
| OpenAI | gpt-4o-mini | 2024-07-18 | 128K tokens |
| Azure OpenAI | gpt-4o-mini | 2024-07-18 | 128K tokens |
| Anthropic (hosted by Amazon Bedrock) | Claude 3.5 Sonnet | 20241022-v2:0 | 200K tokens |

The accuracy calculation process involves a detailed comparison between the LLM-generated exercise analysis results and the corresponding ground truth data. This comparison enables the identification of both successful matches and discrepancies, providing insights into the system's strengths and areas for improvement in exercise analysis capabilities. The integration of multiple LLM providers through LangChain allows for comparative performance analysis across different model architectures and capabilities.

The automated testing pipeline, implemented through our CI/CD system, continuously evaluates each model's performance metrics, including:

- Exercise classification accuracy
- Duration estimation precision
- Caloric expenditure calculation accuracy

Through this comprehensive evaluation methodology, we can assess the effectiveness of our LLM-based exercise analysis system while identifying potential areas for optimization and enhancement. The multi-model approach provides valuable insights into the relative strengths of different LLM providers, enabling optimal model selection for specific analysis tasks while maintaining system reliability and performance.

## 6.3 Evaluation Results

Our evaluation encompasses three critical aspects of exercise analysis: exercise classification (*is_exercise*), duration estimation, and caloric expenditure prediction. The assessment was conducted across three leading LLM models: OpenAI, Azure OpenAI, and Amazon Bedrock's Claude model.

### 6.3.1 Evaluation Metrics Calculation

The performance metrics were calculated using standard classification evaluation formulas. TP, TN, FP, FN are true positive, true negative, false positive, and false negative, respectively.

$$Accuracy = \frac{[TP] + [TN]}{[TP] + [TN] + [FP] + [FN]}$$

$$Precision = \frac{[TP]}{[TP] + [FP]}$$

$$Recall = \frac{[TP]}{[TP] + [FN]}$$

For duration and calories predictions, we considered predictions within a margin of error of 1 unit as correct classifications.

### 6.3.2 Exercise Classification Accuracy

The exercise classification results demonstrate more than 95% accuracy across all models, as illustrated in Table 9 with precision recall curve as shown in Figure 11:

- Bedrock achieved the highest accuracy at 96.491%, with strong precision (97.507%), recall (98.876%), and AUPRC (0.9513)
- Azure OpenAI showed comparable performance with 95.805% accuracy
- OpenAI maintained a solid performance with 95.412% accuracy

Table 9. *is_exercise* classification accuracy evaluation

| Model Provider | Model Name | Number of data | | | | | Evaluation | | | |
|---|---|---|---|---|---|---|---|---|---|---|
| | | Count | TP | TN | FP | FN | Accuracy (%) | Precision (%) | Recall (%) | AUPRC |
| OpenAI | gpt-4o-mini | 741 | 697 | 10 | 19 | 15 | 95.412 | 97.346 | 97.893 | 0.986 |
| Azure OpenAI | gpt-4o-mini | 739 | 698 | 10 | 19 | 12 | 95.805 | 97.350 | 98.310 | 0.986 |
| Anthropic (hosted by Amazon Bedrock) | Claude 3.5 Sonnet | 741 | 704 | 11 | 18 | 8 | 96.491 | 97.507 | 98.876 | 0.987 |

### 6.3.3 Duration Prediction Analysis

Duration prediction showed strong but slightly lower accuracy metrics than exercise classification, as illustrated in Table 10 with precision recall curve as shown in Figure 11:

- Azure OpenAI led with 88.978% accuracy, 97.701% precision, and AUPRC (0.9063)
- OpenAI followed closely with 88.778% accuracy
- Bedrock showed slightly lower accuracy at 86.774% but maintained high precision

Table 10. Exercise duration accuracy evaluation

| Model Provider | Model Name | Number of data | | | | | Evaluation | | | |
|---|---|---|---|---|---|---|---|---|---|---|
| | | Count | TP | TN | FP | FN | Accuracy (%) | Precision (%) | Recall (%) | AUPRC |

| Model Provider | Model Name | Count | TP | TN | FP | FN | | | | |
|---|---|---|---|---|---|---|---|---|---|---|
| OpenAI | gpt-4o-mini | 499 | 424 | 19 | 10 | 46 | 88.778 | 97.696 | 90.213 | 0.995 |
| Azure OpenAI | gpt-4o-mini | 499 | 425 | 19 | 10 | 45 | 88.978 | 97.701 | 90.426 | 0.995 |
| Anthropic (hosted by Amazon Bedrock) | Claude 3.5 Sonnet | 499 | 416 | 17 | 12 | 54 | 86.774 | 97.196 | 88.511 | 0.994 |

6.3.4 Calorie Expenditure Estimation

Caloric expenditure prediction presented the most challenging aspect, as illustrated in Table 11 with precision recall curve as shown in Figure 11:

- Bedrock demonstrated the best performance with 77.200% accuracy and AUPRC (0.8550)
- Azure OpenAI and OpenAI showed similar performance patterns
- All models maintained exceptionally high precision (>99%)

**Table 11.** Caloric expenditure prediction accuracy evaluation

| Model Provider | Model Name | Number of data | | | | | Evaluation | | | |
|---|---|---|---|---|---|---|---|---|---|---|
| | | Count | TP | TN | FP | FN | Accuracy (%) | Precision (%) | Recall (%) | AUPRC |
| OpenAI | gpt-4o-mini | 500 | 335 | 27 | 2 | 136 | 72.400 | 99.407 | 71.125 | 0.989 |
| Azure OpenAI | gpt-4o-mini | 500 | 338 | 27 | 2 | 133 | 73.000 | 99.412 | 71.762 | 0.989 |
| Anthropic (hosted by Amazon Bedrock) | Claude 3.5 Sonnet | 500 | 359 | 27 | 2 | 112 | 77.200 | 99.446 | 76.221 | 0.990 |

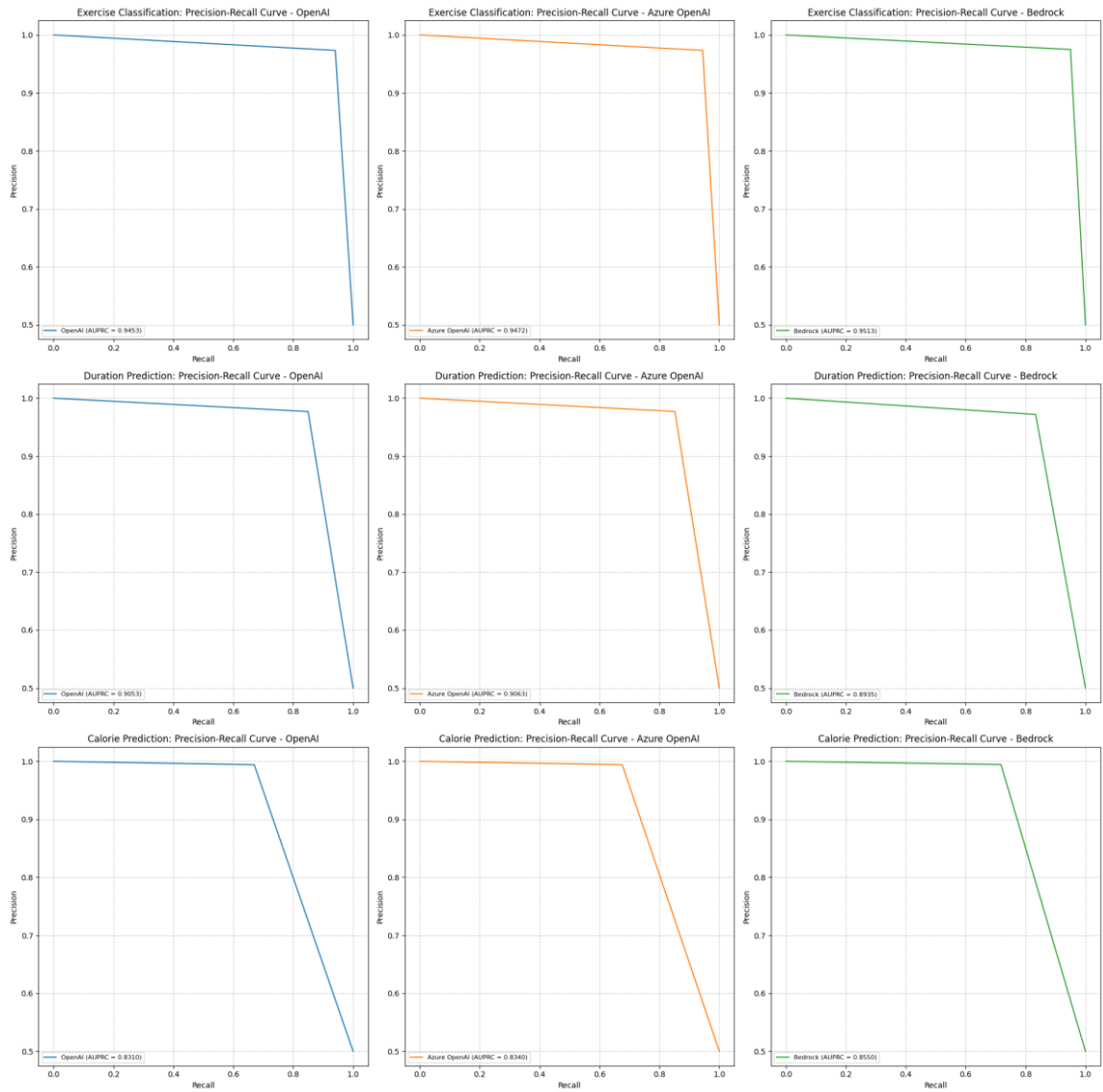

**Figure 11.** Precision recall curve with AUPRC values

6.3.5 Comparative Analysis

The evaluation results reveal key insights through comparative analysis:

- Exercise classification shows consistently high performance across all models

- Duration prediction maintains strong accuracy with minimal variation between models

- Caloric expenditure estimation, while challenging, shows promising results with Bedrock leading in accuracy

These results demonstrate the effectiveness of our LLM-based approach while highlighting areas for potential improvement, particularly in caloric expenditure estimation. The high precision across all metrics suggests reliable performance when positive predictions are made.

## 6.4 Discussion

The evaluation results reveal several interesting challenges and insights in applying LLM-based analysis to exercise-related posts to the community. While exercise classification achieved high accuracy across all models, the prediction of exercise duration and caloric expenditure presented notable challenges due to the diverse nature of user-generated content.

### 6.4.1 Data Quality Variations

The system encountered varying levels of information completeness in user posts. Exercise classification proved relatively straightforward due to clear contextual indicators, but duration and calorie calculations faced several challenges:

- Insufficient context in posts where users only shared screenshots without descriptive text
- Lack of standardized format for exercise information presentation
- Variable quality and completeness of exercise documentation

### 6.4.2 Multiple Exercise Scenarios

The analysis of posts containing multiple exercise activities presented unique challenges:

- Posts combining different types of exercises in a single update
- Sequential posts of individual exercises from the same session
- While exercise identification remained accurate, calculating total duration and calories became more complex in these scenarios

### 6.4.3 Temporal Aggregation Issues

Some users' posting behaviors created temporal analysis challenges:

- Multiple-day exercise summaries posted in single updates
- Deviation from the community's typical daily posting pattern
- These variations complicated the LLM's ability to accurately parse temporal exercise data

### 6.4.4 Measurement Precision Challenges

Technical aspects of measurement created additional complexity:

- Rounding discrepancies in minute-level duration calculations
- 1-2 minute variations in total duration calculations due to second-level precision differences
- Inconsistencies in how different exercise tracking devices report time measurements

### 6.4.5 Caloric Measurement Ambiguity

A significant challenge emerged in caloric expenditure calculation:

- Confusion between total calories and active calories in exercise reports
- Inconsistent reporting of caloric metrics across different exercise tracking platforms

- LLM uncertainty in selecting appropriate caloric values for aggregation

6.4.6 Future Improvements

The identified challenges suggest several areas for potential enhancement:

- Refined prompt engineering techniques to better handle caloric measurement ambiguity
- Implementation of standardized rules for temporal data aggregation
- Development of more sophisticated multiple-exercise analysis algorithms

# 7 Conclusion

This paper presents an innovative approach to exercise analysis and feedback generation through the integration of LLMOps in a social healthcare platform, specifically focusing on the Ounwan exercise community. The research explores the potential of leveraging LLMOps to provide automated, personalized feedback and analysis of user activity records, demonstrating significant improvements in healthcare community management.

The system successfully implements automated exercise data collection, analysis, and personalized feedback generation through sophisticated LLM integration. Through comprehensive evaluation using multiple LLM providers (OpenAI, Azure OpenAI, and Amazon Bedrock), the system demonstrated robust performance in exercise classification (>95% accuracy), duration prediction (>86% accuracy), and caloric expenditure estimation (>72% accuracy).

The research emphasizes the effectiveness of LLMOps in improving the efficiency and reliability of large-scale machine learning models, particularly in driving personalized recommendations that align closely with user preferences. The evaluation results demonstrate the system's capability to enhance physical activity levels while potentially reducing the risk of chronic diseases in individuals with sedentary lifestyles.

A key focus of the implementation has been on ethical considerations and data security. The system architecture prioritizes model interpretability and transparency, ensuring that automated feedback maintains high accuracy while protecting user privacy. The integration of multiple LLM providers through a containerized infrastructure demonstrates the scalability and reliability of the approach, while maintaining strict data security standards essential for healthcare applications.

The paper provides comprehensive evidence for the feasibility and effectiveness of AI-powered health interventions in community-based platforms, particularly through the systematic application of LLMOps in exercise analysis and feedback generation.

**Acknowledgements:** We sincerely thank the organizers of the Ounwan social community (https://www.facebook.com/groups/1318095768809442) for allowing us to utilize their publicly available data for this study. Their generosity and collaboration have greatly contributed to this research.